# Phenomenon of irreducible genetic markers for TATAAA motifs in human chromosome 1

Sergey V. Chesnokov[1]* · Lina G. Chesnokov[1] · Viktor Wixler[2]

[1] Department of Mathematics, Bar Ilan University, Ramat Gan 52900, Israel.

[2] Institute of Molecular Virology, Centre of Molecular Biology of Inflammation, University of Muenster, Von-Esmarch-Str. 56, D-48149 Muenster, Germany.

* corresponding author, e-mail: sergeyches@gmail.com

**Abstract** In searching for rules that might allow identification and localization of distinct motifs within large sequence chains of DNA or proteins, similar to language orthography which enables the efficient sounding of letters when reading in English, we analyzed the "orthography" of the canonical Goldberg-Hogness DNA transcriptional start site, the TATAAA motif. Employing the principles of determinacy analysis (mathematical theory of rules) along with a fragment of human chromosome 1 DNA sequence, we identified and described specific genetic markers present in the nearest proximity to TATAAA motifs that enable determining the exact location of any TATAAA motif in the DNA fragment. Individual, as well as groups of TATAAA motifs could be identified. We termed these markers irreducible genetic (IG) markers, as all bases are essential, and removal of any base abolishes their function. In the DNA fragment that was examined, we found and analyzed all IG-markers for every TATAAA motif, regardless of its role in gene transcription. It is well known that the general transcription factors (GTF) specifically recognize correct TATA boxes, distinguishing them from many others. Our results led us to hypothesize that the GTF recognize the «true» transcriptional start TATA box by means of IG-markers. The mathematical method described here is universal and can be used to find IG-markers that will provide like a global navigation satellite system for the specific location of any distinct sequence motif within larger DNA sequence content.

**Abbreviations** IG-marker = irreducible genetic marker · IG-marker o.c. = IG-marker occurrence case · TBP = TATA binding protein · GTF = general transcription factors

---

**Funding** Emmy Noether Research Institute for Mathematics, Bar-Ilan University, Israel; Department of Mathematics, Bar-Ilan University, Israel; Israel Scientific Foundation grant 688/08, partial support.

# 1 Introduction

We were wondering whether "orthography rules" are hidden in arrays of DNA sequences enabling their correct “reading” by enzymes, similar to orthography rules that are hidden in arrays of English text, which ensure the correct sounding of printed words when reading (Chesnokov and Luelsdorff 1991; Luelsdorff and Chesnokov 1994, 1996). To explore the issue, we conducted a study of "orthography rules" that bind or integrate any TATAAA motif into its adjacent DNA sequence. We employed the computational algorithms of determinacy analysis, which is the mathematical theory of rules that are hidden in observational data or text arrays in alphabetic orthographies (Chesnokov 1982, 1987, 1990, 2009). The criteria underlying the idea of orthography rules for any TATAAA motif were previously applied to study protein orthography (Chesnokov and Reznik 2002, unpublished).

There were three reasons to choose TATAAA motifs.

1) The TATAAA site is the classical Goldberg-Hogness TATA box.
2) The transcription of 10 to 15 % of all mammalian genes are initiated at TATA boxes, located 25 to 35 bp upstream of the transcriptional start site (Lifton et al 1978; Woychik and Hampsey 2002; Cooper et al 2006; Juven-Gershon and Kadonaga 2010).
3) The TATAAA motif is the recognition site for the TATA binding protein (TBP), which is part of the general transcription factor protein complex. After TBP recognizes and binds to the TATAAA, the general transcription factors (GTF) assemble at promoter utilized by RNA polymerase II, and then the transcription can start.

Our considerations were the following. It is well known that there are many more TATAAA motifs in the genome sequence than transcribed genes. Alone the DNA sequence of the human chromosome 1, ftp://ftp.ensembl.org/pub/release-57/fasta/homo_sapiens/dna/, contains 151656 TATAAA motives. It is much more than the approximated maximal number of genes in the total human genome. Hence the question arises as to how the GTF recognize the "true" TATAAA box, distinguishing it from many other TATAAA motifs? Examination of the DNA as a simple text sequence only did not permit an answer to this question.

However, by applying method described in this work, we were able to identify combinations of bases located in the close proximity to TATAAA motifs that can serve as potential landmarks (reference points) for GTF, allowing them to choose exactly the "true" TATAAA box. We termed them *irreducible genetic markers* or *IG-markers*. Whether GTF really use IG-markers as the landmarks, it requires separate experimental verification.

# 2 Methods

## 2.1 What is an IG-marker?

Any IG-marker for a TATAAA motif group, if it exists in investigated DNA fragments, is defined by the following characteristics, which should be valid in these DNA fragments:

- An IG-marker for a TATAAA motif group provides recognition for the DNA position of every TATAAA motif of that group, and does it with the probability equals 1.
- All bases of an IG-marker are essential (irreducible). No base can be removed from an IG-marker, hence the "Irreducible" in the term "IG-marker".

From the standpoint of determinacy analysis, the mathematical meaning of these characteristics is given by equation (7) (see below precise definition of an IG-marker) and this is the basis for the calculation algorithm applied here for detection and analysis of IG-markers. In more details, suppose *B* is the TATAAA motif with its particular position in DNA sequence when set of bases *A* is the IG-marker for *B*, and *A′* is the complement of any base $X \in A$ in the set *A*. In terms of DNA orthography, the IG-marker *A* is the argument of the orthography rule "*If A, then B*", when this is true with conditional probability $P(B|A) = 1$, such that $P(B|A') < 1$. Now, if we rewrite *B*, *A*, and *A′* in these two restrictions, taking in account some other details, we obtain IG-marker equation (7) (see below).

There are some additional points to be considered. From the molecular genetics point of view, IG-markers are the partial case of ordinal genetic markers. Regarding the theoretical linguistics for alphabetic languages, IG-markers are sets of binders that provide the governing and binding of functional sites in arrays of DNA text. In alphabetic orthographies, the idea of government by means of binders was proposed by Chomsky (1981).

DNA fragments studied in this work contain 128 TATAAA motifs. For every individual TATAAA motif, the 12 bp long sequence downstream of the TATAAA motif (5'→3' direction) was analyzed. Many of the general transcription factors are concentrated in this area, as TATAAA motif is the transcriptional start TATA box. The primary task of this work, however, was to resolve the phenomenon of IG-markers for TATAA motifs in general. Therefore we studied here all IG-markers in this 12 bp sequence that bind the TATAAA motif to its particular sequence, irrespective of whether this TATAA motif is localized in promoter regions or not. The sequence length of 12-bases was recognized to be sufficient to resolve the phenomenon still occupying a rational computational time. Whether IG-markers might be used as landmarks for GTF to define the "true" TATA box is certainly an issue for future studies of DNA orthography.

## 2.2 Data Base (DB1)

We used the DNA sequence of human chromosome 1 'Homo_sapiens.GRCh37.57.dna.chromosome.1.fa' published on 3 Mar 2010 (see ftp://ftp.ensembl.org/pub/release-57/fasta/homo_sapiens/dna/). This sequence contains a total of 249 250 621 bp; of these, 225 280 621 bp have already been sequenced composed of 38 contig regions including 151 656 TATAAA motifs.

The analyzed database (DB1) consist of the 5 (of 38) shortest contig regions. In DNA sequence of human chromosome 1 their ordinal numbers are 2, 7, 21, 24 and 27 (in the 5'→3' orientation). Total length of DB1 DNA sequence is 312874 bp. The total number of TATAAA motifs is 128 (for details see Table 1).

**Table 1.** Contig regions of DB1 and their characteristics

| (1) | (2) | (3) | (4) |
|---|---|---|---|
| 2 | 227418 | 40302 | 44 |
| 7 | 13103000 | 116914 | 38 |
| 21 | 144145784 | 78698 | 20 |
| 24 | 144672414 | 38311 | 22 |
| 27 | 146214651 | 38649 | 4 |
| | Total: | 312874 | 128 |

Column (1): ordinal number of contigs. Column (2): start position of contig regions in the DNA sequence of the human chromosome 1. Column (3): length of contigs in bp. Column (4): number of TATAAA motifs in contigs.

The results of this work are also valid in the same set of contigs in the last updated sequence of the human genome, version 'Homo_sapiens.GRCh37.-62.dna.chromosome.1.fa' published on 12 Apr 2011 (see ftp://ftp.ensembl.org/pub/release-62/fasta/homo_sapiens/dna/).

## 2.3 Problem, mathematics and calculations

### 2.3.1 TATAAA motifs and accompanying definitions

By TATAAA[*w*] we denote a single TATAAA motif, where the last (from left to right) A occupies position *w* in the DNA sequence counted from 5' to 3'. We say that the integer *w* is, by definition, «the position of TATAAA[*w*] in the DNA sequence».

Let $\Im_s(w)$ be a sequence of length *s* (set of bases) that follows TATAAA[*w*]:

$$\Im_s(w) = \{X_{w+1}, X_{w+2}, ..., X_{w+s}\}, \qquad (1)$$

where $X_{w+k}$ is a base of A, T, G or C at position $w+k$ in DNA, $k = 1,2,...,s$. In DNA, TATAAA[*w*] and sequence $\Im_s(w)$ form the configuration

$$\text{TATAAA}[w]\Im_s(w) = \text{TATAAA}[w]X_{w+1}X_{w+2}...X_{w+s}. \qquad (2)$$

Let $\Im_s(w,r)$ be a set of *r* bases that is a subset of the set $\Im_s(w)$, $1 \le r \le s$:

$$\Im_s(w,r) = \{X_{w+q_1}, X_{w+q_2}, ..., X_{w+q_r}\} \subseteq \Im_s(w). \qquad (3)$$

Here $1 \le q_1 < q_2 < ... < q_r \le s$.

Suppose, for a DNA position $z \ne w$ there exists a set of bases

$$\Im_s(z,r) = \{X_{z+q_1}, X_{z+q_2}, ..., X_{z+q_r}\}, \qquad (4)$$

where $X_{z+q_i} = X_{w+q_i} \in \Im_s(w,r)$, $i = 1,2,...,r$. This means that the bases of the set $\Im_s(z,r)$ and their relative locations in DNA are the same as in the set $\Im_s(w,r)$. In other words, the set $\Im_s(z,r)$ can be considered as a result of shift $\Im_s(w,r) \to \Im_s(z,r)$ along DNA sequence whenever $w \to z$. Thus, in this special case, we say that *z* and *w* *are the DNA addresses for* $\Im_s(w,r)$ and $\Im_s(z,r)$ is said to be *the occurrence case (o.c.) of* $\Im_s(w,r)$ in DNA sequence.

The set containing all DNA addresses for the set $\Im_s(w,r)$ is called *an address set for* $\Im_s(w,r)$ *in DNA* and is denoted by $G_w$:

$$G_w = \{z \mid X_{z+q_i} = X_{w+q_i}; i = 1,2,...,s\}. \qquad (5)$$

The address set $G_w$ for $\Im_s(w,r)$ and the set $\{\Im_s(z,r) \mid z \in G_w\}$ are uniquely determined by the set $\Im_s(w,r)$ and by analyzed DNA text (DB1, in this work).

### 2.3.2 IG-marker for TATAAA motif group

Denote by $\aleph_v$ any TATAAA motif group[1] containing $v$ motifs of TATAAA[$z$], $z \in H$ in analyzed DNA text:

$$\aleph_v = \{\mathrm{TATAAA}[z] \mid z \in H\},\ v = |H| = |\aleph_v|. \tag{6}$$

Consider the following problem. For some $z = w \in H$, now assume that the $\mathrm{TATAAA}[w] \in \aleph_v$, the set of bases $\Im_s(w)$ and the set $\Im_s(w,r) \subseteq \Im_s(w)$ are given (see (1), (2), (3)). For analyzed DNA text, the question arise whether $\Im_s(w,r)$ is IG-marker for $\aleph_v$ or not? The answer is provided by the IG-marker definition.

*IG-marker definition*. Let $G_w$ be the address set for $\Im_s(w,r)$. The set $\Im_s(w,r)$ is said to be an *IG-marker* with rank $r$ for TATAAA motif group $\aleph_v$ (and can be used as identifier for every TATAAA motif of $\aleph_v$) whenever $H = G_w$ such that for every $z \in G_w$ the set $\Im_s(z,r)$ satisfies the equation

$$\begin{cases} \mathbf{P}(\mathrm{TATAAA}[z] \in \aleph_v \mid \Im_v(z,r)) = 1, \\ 1 - \mathbf{P}(\mathrm{TATAAA}[z] \in \aleph_v \mid \Im_s(z,r;\overline{\mathrm{X}}_{q_i})) > 0,\ i = 1,2,\ldots,r. \end{cases} \tag{7}$$

Here in the top row, $\mathbf{P}(B \mid A)$ is the conditional probability of event $B$ = ( $\mathrm{TATAAA}[z] \in \aleph_v$) given event $A$ = $\Im_s(z,r)$. In the second row, $\mathbf{P}(B \mid A')$ is the conditional probability of event $B$ given event $A'$ = $\Im_s(z,r;\overline{\mathrm{X}}_{q_i})$ that is
a complement of base $\mathrm{X}_{z+q_i}$ in the set $\Im_s(z,r)$:

$$\Im_s(z,r;\overline{\mathrm{X}}_{q_i}) = \{\mathrm{X}_{z+q_1},\ldots,\mathrm{X}_{z+q_{i-1}},\mathrm{X}_{z+q_{i+1}},\ldots,\mathrm{X}_{z+q_r}\} = \Im_s(z,r) \setminus \mathrm{X}_{q_i}. \tag{8}$$

In DNA sequence, the $\Im_s(z,r)$ is called *the occurrence case (o.c.) of IG-marker* $\Im_s(w,r)$ if $z \in G_w$ (see below (13) and explanation to it).

### 2.3.3 IG-marker equation

The (7) is called the *IG-marker equation*. For every $z \in G_w$ the equality in the top row of (7) means the rule (9) is true with the probability 1:

$$\text{If } \Im_s(z,r), \text{ then } \mathrm{TATAAA}[z] \in \aleph_v. \tag{9}$$

---

[1] Here and subsequently, the word "group" is only synonym to "set".

The second row in (7) contains the strict inequalities for $r$ differences

$$1-\mathbf{P}(\mathrm{TATAAA}[z]\in\aleph_v \mid \Im_s(z,r;\overline{\mathrm{X}}_{q_i}))\equiv\Delta\mathbf{P}(\mathrm{X}_{q_i}),\ i=1,2,...,r. \qquad (10)$$

We use the $\Delta\mathbf{P}(\mathrm{X}_{q_i})$ value as the measure of essentiality (irreplaceability) of the base $\mathrm{X}_{z+q_i}$ in the set $\Im_s(z,r)$. If $\Delta\mathbf{P}(\mathrm{X}_{q_i})=0$, then the base $\mathrm{X}_{z+q_i}$ is *nonessential* (*replaceable, reducible*). Thus the strict inequalities in (7) require every base of an IG-marker must be *essential* (*irreplaceable, irreducible*):

$$\Delta\mathbf{P}(\mathrm{X}_{q_i})>0, i=1,2,...,r. \qquad (11)$$

Without inequalities (11) the equation (7) defines an *ordinary G-marker* (*genetic marker*) for the TATAAA motif group $\aleph_v$.

### 2.3.4 Some general properties of IG-markers

The following properties of IG-markers for a TATAAA motif group $\aleph_v$ can be derived from the equation (7):

1) Any IG-marker is a special case of ordinary G-markers. IG-markers cannot contain unessential (reducible) bases (see (10) and (11)), whereas an ordinary G-markers might have such bases.
2) Denote by $G^*$*-marker* any ordinary G-marker that is not an IG-marker. If $E_v^*$ is the total set of all $G^*$-markers for TATAAA group $\aleph_v$, then every $G^*$-marker from the set $E_v^*$ contains a non-empty set of IG-markers for $\aleph_v$.
3) A $G^*$-marker is a *minimal* $G^*$*-marker in the set* $E_v^*$ if it contains a minimal number of bases among all $G^*$-markers in $E_v^*$. If $s_{\min}^*$ is the number of bases in a minimal $G^*$-marker in $E_v^*$ and $r_{\max}$ is the maximal rank among all IG-markers that are included in the minimal $G^*$-marker, then in $E_v^*$ the strict inequality holds:

$$r_{\max}<s_{\min}^*. \qquad (12)$$

4) IG-marker $\Im_s(w,r)$ for TATAAA motif group $\aleph_v$ having volume $v$ has $v$ occurrence cases (o.c.'s) $\Im_s(z,r)$, $z\in G_w$ in the DNA sequence:

1 IG-marker = $v$ IG-marker o.c.'s (13)

5) Rank of an IG-marker (the number of bases in it) represents a local optimum. If any base is removed from an IG-marker for the TATAAA motif group $\aleph_v$, then the remaining set of bases is not a genetic marker for $\aleph_v$. If any new base is added to any IG-marker for the TATAAA motif group $\aleph_v$, then the new set of bases or becomes a $G^*$-marker for $\aleph_v$, or it is not a genetic marker for $\aleph_v$ at all.

6) Consider two IG-markers $\Im_s(w,r)$, $\Im_s(w,r')$ for a TATAAA motif group $\aleph_v$ as $r < r'$. It is impossible to transform $\Im_s(w,r) \to \Im_s(w,r')$ by only adding bases to $\Im_s(w,r)$ or only removing bases from $\Im_s(w,r')$.
7) Within an IG-marker, adjacent bases are not required to be nearest neighbors in the DNA sequence.

### *2.3.5 The TATAAA motif problem*

DB1 contains 128 $\mathrm{TATAAA}[w_t]\Im_s(w_t)$ configurations (see (1), (2)):

$$\mathrm{TATAAA}[w_t]\{\mathrm{X}_{w_t+1}, \mathrm{X}_{w_t+2}, ..., \mathrm{X}_{w_t+s}\},\ t = 1,2,...,128. \qquad (14)$$

For $s = 12$, Table 3 shows all of these 128 configurations and their positions $w_t, t = 1,2,...,128$ in DNA of human chromosome 1 (columns '$t$', '$w_t$' and '$\mathrm{TATAAA}[w_t]\mathrm{XXXXXXXXXXXX}$'). The two following problems to be solved:

*Problem 1*. For $s = 12$ (see Introduction) and for every $t = 1,2,...,128$, the complete set of IG-markers for $\mathrm{TATAAA}[w_t]$ located in the set of bases $\Im_s(w_t) = \{\mathrm{X}_{w_t+1}, \mathrm{X}_{w_t+2}, ..., \mathrm{X}_{w_t+s}\}$ (see (14)) must be found.

*Problem 2*. To find all TATAAA motif groups $\aleph_v$, $v \geq 1$ that are formed by IG-markers which were found as the solution of problem 1.

### *2.3.6 Algorithm to solve problems 1 and 2*

Let any TATAAA[$w_t$] with the following it sequence $\Im_s(w_t)$ (see (14)) be fixed. Consider set $\Im_s(w_t,r) \subseteq \Im_s(w_t)$, its address set $G_{w_t}$ and sets $\Im_s(z_t,r), z_t \in G_{w_t}$:

$$\Im_s(w_t,r) = \{\mathrm{X}_{w_t+q_1}, \mathrm{X}_{w_t+q_2}, ..., \mathrm{X}_{w_t+q_r}\} \subseteq \Im_s(w_t),\ 1 \leq r \leq s; \qquad (15)$$

$$G_{w_t} = \{z_t \mid \mathrm{X}_{z_t+q_t} = \mathrm{X}_{w_t+q_t}; i = 1,2,...,r\}; \qquad (16)$$

$$\Im_s(z_t,r) = \{\mathrm{X}_{z_t+q_1}, \mathrm{X}_{z_t+q_2}, ..., \mathrm{X}_{z_t+q_r}\}, z_t \in G_{w_t}. \qquad (17)$$

From IG-marker definition (see (7)) it follows that the set $\Im_s(w_t,r)$ is the IG-marker of rank $r$ for TATAAA motif group $\aleph_v$ and $\mathrm{TATAAA}[w_t] \in \aleph_v$ if the set $\Im_s(z_t,r)$ satisfies equation (7) for every $z_t \in G_{w_t}$.

If rank $r$ is not fixed in the range $1 \leq r \leq s$, then the quantity $\Re(s)$ of combinatorial variants for the non-empty set $\Im_s(w_t,r)$ is equal to $2^s - 1$. If $r$ is fixed, then quantity $\Re(s;r)$ of combinatorial variants for the set $\Im_s(w_t,r)$ is given by the binomial coefficient in series

$$\Re(s) = \sum_{r=1}^{r=s} \Re(s;r) = \sum_{r=1}^{r=s} \frac{s!}{r!(s-r)!} = 2^s - 1 \,. \tag{18}$$

To solve problems 1 and 2, all of $2^s - 1$ combinatorial variants of the set $\Im_s(w_t, r)$, $r$ = 1,2,…,$s$ must be surveyed for every $t$ = 1,2,…,128. For every variant of $\Im_s(w_t, r)$ the sets $\Im_s(z_t, r)$, $z_t \in G_{w_t}$ must be found. And if all sets $\Im_s(z_t, r)$, $z_t \in G_{w_t}$ ($t$, $r$ are fixed) satisfy equation (7), the set $\Im_s(w_t, r)$ must be selected as the IG-marker for TATAAA motif group $\aleph_v$. The C++ version of this algorithm was applied here to solve problems 1 and 2.

# 3 Results and discussions

## 3.1 General set of IG-markers in DB1

In sequences (sets) $\Im_s(w_t)$ of length $s$ = 12, $t$ = 1,2,…,128 (see (14)), there are 4765 IG-markers for TATAAA motif groups $\aleph_v$ having volume $v$ = 1, 2, 16: 4630 markers for $v$ = 1; 115 markers for $v$ = 2, and 20 markers for $v$ = 16. In accordance with equation (13), the total number of IG-marker o.c.'s is equal to 5180, with 4630 o.c.'s being valid for $v$ = 1, 230 o.c.'s for $v$ = 2, and 320 o.c.'s for $v$ = 16.

Table 2 shows the distribution of IG-marker o.c. numbers in analyzed contig regions and all TATAAA motif groups having volume $v$ = 1, 2 and 16.

**Table 2.** Distribution of 5180 IG-marker o.c.'s in analyzed contig regions and TATAAA motif groups

| Contigs | $v$ = 1 | $v$ = 2 | $v$ = 16 | Total |
|---|---|---|---|---|
| 2 | 2301 | 1 | 0 | 2302 |
| 7 | 1374 | 66 | 0 | 1440 |
| 21 | 105 | 5 | 320 | 430 |
| 24 | 850 | 8 | 0 | 858 |
| 27 | 0 | 150 | 0 | 150 |
| Total: | 4630 | 230 | 320 | 5180 |

Column 'Contigs': ordinal number of contig regions. Column '$v$ = 1': IG-marker o.c. numbers for all TATAAA motif groups having volume $v$ = 1. Column '$v$ = 2': IG-marker o.c. numbers for all TATAAA motif groups having volume $v$ = 2; Column '$v$ = 16': IG-marker o.c. numbers for all TATAAA motif groups having volume $v$ = 16.

Table 3 (see below, columns 'v = 1', 'v = 2', 'v = 16') shows the distribution of all 5180 identified IG-marker o.c. numbers among TATAAA[$w_t$], $t$ = 1,2,…,128.

**Table 3.** IG-marker o.c. numbers for every TATAAA[$w_t$], $t$ = 1,2,…,128 in DB1 (five studied contig regions of the human chromosome 1)

***Contig 2.*** In DNA from 227418 bp to 267719 bp; 44 TATAAA motifs; IG-marker o.c.'s total: 2302 (2301 for $v$ = 1 and 1 for $v$ = 2).

| $t$ | $w_t$ (bp) | TATAAA[$w_t$] XXXXX XXXXX | $v$=1 | $v$=2 | $v$=16 | $s^*_{min}$ | $t$ | $w_t$ (bp) | TATAAA[$w_t$] XXXXX XXXXX | $v$=1 | $v$=2 | $v$=16 | $s^*_{min}$ |
|---|---|---|---|---|---|---|---|---|---|---|---|---|---|
| 1 | 227973 | TATAAA TAAAT ATTCA CC | 27 | 0 | 0 | 11 | 23 | 252216 | TATAAA AGTGT TGACT TG | 50 | 0 | 0 | 10 |
| 2 | 228126 | TATAAA ACAAG CATAA AT | 64 | 0 | 0 | 9 | 24 | 253076 | TATAAA CCTCC AGAAT AA | 49 | 0 | 0 | 10 |
| 3 | 228924 | TATAAA TAAGA TACAA TA | 45 | 0 | 0 | 9 | 25 | 253614 | TATAAA GGAGT GGAAA AC | 68 | 0 | 0 | 9 |
| 4 | 229432 | TATAAA ATTCC TGGAA GT | 31 | 0 | 0 | 11 | 26 | 255297 | TATAAA ACAGG AACAT CT | 33 | 0 | 0 | 10 |
| 5 | 229652 | TATAAA AAATT AGCCG GG | 0 | 0 | 0 | 17 | 27 | 256781 | TATAAA CCATT CTACC AT | 58 | 0 | 0 | 10 |
| 6 | 231086 | TATAAA TTACG CAGCC TC | 93 | 0 | 0 | 8 | 28 | 257906 | TATAAA GTATA AATTT GT | 30 | 0 | 0 | 10 |
| 7 | 231225 | TATAAA GTTCT TTCCA AC | 37 | 0 | 0 | 10 | 29 | 257913 | TATAAA TTTGT GATTT TG | 25 | 0 | 0 | 12 |
| 8 | 232634 | TATAAA TCAAA ACAAA TG | 0 | 0 | 0 | 14 | 30 | 258131 | TATAAA ATACC GAGAT TA | 101 | 0 | 0 | 8 |
| 9 | 233674 | TATAAA ACCAA GCTGT GC | 0 | 0 | 0 | 16 | 31 | 258154 | TATAAA CAACT TTAGA TT | 59 | 0 | 0 | 11 |
| 10 | 233841 | TATAAA AGGGA TGTTA TT | 42 | 0 | 0 | 11 | 32 | 258331 | TATAAA ATTAC TGTTT AG | 60 | 0 | 0 | 10 |
| 11 | 242039 | TATAAA GAACC CAGCG TG | 72 | 0 | 0 | 10 | 33 | 261363 | TATAAA ACAGA CCTCT TC | 66 | 0 | 0 | 10 |
| 12 | 242577 | TATAAA CTGCC AATCA TG* | 62 | 1 | 0 | 12 | 34 | 261525 | TATAAA GAATT GTCCA GA | 48 | 0 | 0 | 12 |
| 13 | 242948 | TATAAA AAACG GCCTT TT | 79 | 0 | 0 | 8 | 35 | 261547 | TATAAA AAAAG AATCA CA | 10 | 0 | 0 | 10 |
| 14 | 243468 | TATAAA AAGGT GAGCT GT | 76 | 0 | 0 | 9 | 36 | 263049 | TATAAA CATAG AGATT GC | 69 | 0 | 0 | 10 |
| 15 | 243672 | TATAAA CATCA GCCAA GT | 64 | 0 | 0 | 10 | 37 | 263566 | TATAAA GCACT AACCA TT | 96 | 0 | 0 | 9 |
| 16 | 244115 | TATAAA CATGT AGCAT TG | 70 | 0 | 0 | 9 | 38 | 264617 | TATAAA GCTTC TCTAT TT | 34 | 0 | 0 | 10 |
| 17 | 245154 | TATAAA CAAAT CTGTC TA | 68 | 0 | 0 | 10 | 39 | 264972 | TATAAA CTGAT ACAGC TA | 71 | 0 | 0 | 10 |
| 18 | 246529 | TATAAA AATAT GCCTC AG | 0 | 0 | 0 | 15 | 40 | 265770 | TATAAA TATAT ATGTA CA | 42 | 0 | 0 | 11 |
| 19 | 248486 | TATAAA TCAAA GCACT AT | 46 | 0 | 0 | 9 | 41 | 265797 | TATAAA ATACT AAACA AA | 43 | 0 | 0 | 9 |
| 20 | 248868 | TATAAA GCCAC CGTTT AT | 111 | 0 | 0 | 9 | 42 | 265914 | TATAAA GAACT GCCCA AG | 59 | 0 | 0 | 11 |
| 21 | 250859 | TATAAA ACTTG CTAAC AC | 77 | 0 | 0 | 10 | 43 | 265943 | TATAAA GGAAA GAAGT TT | 70 | 0 | 0 | 10 |
| 22 | 251188 | TATAAA CAACA CTGAG CT | 46 | 0 | 0 | 10 | 44 | 266094 | TATAAA ACCAT CAAAT CT | 50 | 0 | 0 | 10 |

***Contig 7.*** In DNA from 13103000 bp to 13219913 bp; 38 TATAAA motifs; IG-marker o.c.'s total: 1440 (1374 for $v$ = 1 and 66 for $v$ = 2).

| $t$ | $w_t$ (bp) | TATAAA[$w_t$] XXXXX XXXXX | $v$=1 | $v$=2 | $v$=16 | $s^*_{min}$ | $t$ | $w_t$ (bp) | TATAAA[$w_t$] XXXXX XXXXX | $v$=1 | $v$=2 | $v$=16 | $s^*_{min}$ |
|---|---|---|---|---|---|---|---|---|---|---|---|---|---|
| 45 | 13104426 | TATAAA TATAT GTATT TC | 32 | 0 | 0 | 11 | 64 | 13168607 | TATAAA AACTA TATAC GA | 102 | 0 | 0 | 10 |
| 46 | 13110507 | TATAAA AATTA GCCAG GT | 0 | 0 | 0 | 21 | 65 | 13175448 | TATAAA AGACA GATAT AG | 52 | 0 | 0 | 10 |
| 47 | 13118494 | TATAAA AATGC TTTCT AC | 28 | 0 | 0 | 12 | 66 | 13176380 | TATAAA ATTGT TTTAT AG | 24 | 0 | 0 | 11 |
| 48 | 13118524 | TATAAA AACAG AGGAG TC* | 34 | 19 | 0 | 10 | 67 | 13176608 | TATAAA TTTAA TCGGT GA | 82 | 0 | 0 | 8 |
| 49 | 13118723 | TATAAA TTTAA TCAGT GA | 0 | 0 | 0 | 24 | 68 | 13178060 | TATAAA AGCAT TCCAA CT | 0 | 0 | 0 | 33 |
| 50 | 13122319 | TATAAA TGGCT AAAAC AG | 46 | 0 | 0 | 10 | 69 | 13178261 | TATAAA GTAAT CCTTG CA* | 0 | 12 | 0 | 21 |
| 51 | 13123779 | TATAAA GGCCC CCAAG AG | 43 | 0 | 0 | 9 | 70 | 13178926 | TATAAA ATACA GGTTT GA | 44 | 0 | 0 | 10 |
| 52 | 13124876 | TATAAA GCATC TGAGG GT | 55 | 0 | 0 | 10 | 71 | 13181521 | TATAAA ACCCA ATTGT AG | 80 | 0 | 0 | 9 |
| 53 | 13127284 | TATAAA AAAAT TTAAT AG | 32 | 0 | 0 | 10 | 72 | 13182023 | TATAAA AATGA GCTGC TC | 48 | 0 | 0 | 10 |
| 54 | 13127682 | TATAAA CTGCA AATTA AG* | 42 | 1 | 0 | 11 | 73 | 13185071 | TATAAA ATTTA TAAAA AT | 0 | 0 | 0 | 15 |
| 55 | 13129277 | TATAAA ATGGC CCCAC CC | 51 | 0 | 0 | 10 | 74 | 13185080 | TATAAA AATAT TTCTA GT | 45 | 0 | 0 | 11 |
| 56 | 13132426 | TATAAA ATTAG AACAT GA | 29 | 0 | 0 | 10 | 75 | 13185109 | TATAAA CACAG AGGAG TC* | 35 | 19 | 0 | 11 |
| 57 | 13136464 | TATAAA GTTTA AACTC TG | 30 | 0 | 0 | 10 | 76 | 13186954 | TATAAA GTAAT CCTTG CA* | 0 | 12 | 0 | 21 |
| 58 | 13143911 | TATAAA TGATG GTGAA AT | 33 | 0 | 0 | 12 | 77 | 13187392 | TATAAA TGTTA ACCAA CC | 53 | 0 | 0 | 10 |
| 59 | 13149708 | TATAAA CATCC AGGTG TG | 35 | 0 | 0 | 10 | 78 | 13188822 | TATAAA TATAT AGAAA GT* | 45 | 3 | 0 | 10 |
| 60 | 13152629 | TATAAA AATCC TGCTC AG | 37 | 0 | 0 | 10 | 79 | 13206855 | TATAAA TACGC CACTG GG | 114 | 0 | 0 | 7 |
| 61 | 13162119 | TATAAA TCCCT AAAAC AA | 47 | 0 | 0 | 10 | 80 | 13212578 | TATAAA GAAAA AAATA GG | 30 | 0 | 0 | 11 |
| 62 | 13166080 | TATAAA AAAAT TAAAA AT | 5 | 0 | 0 | 12 | 81 | 13218358 | TATAAA AATTA GCCAG GT | 0 | 0 | 0 | 21 |
| 63 | 13166562 | TATAAA AGCAG TGTTG CA | 41 | 0 | 0 | 10 | 82 | 13219051 | TATAAA GCTTT TGGAG GC | 0 | 0 | 0 | 15 |

***Contig 21.*** In DNA from 144145784 bp to 144224481 bp; 20 TATAAA motifs; IG-marker o.c.'s total: 430 (105 for $v$ = 1; 5 for $v$ = 2 and 320 for $v$ = 16)

| $t$ | $w_t$ (bp) | TATAAA[$w_t$] XXXXX XXXXX | $v$=1 | $v$=2 | $v$=16 | $s^*_{min}$ | $t$ | $w_t$ (bp) | TATAAA[$w_t$] XXXXX XXXXX | $v$ = 1 | $v$=2 | $v$=16 | $s^*_{min}$ |
|---|---|---|---|---|---|---|---|---|---|---|---|---|---|
| 83 | 144149736 | TATAAA GTCCT GGTTC AC | 52 | 0 | 0 | 9 | 93 | 144186052 | TATAAA GATCA TATTC CT** | 0 | 0 | 20 | >100 |
| 84 | 144151800 | TATAAA CACAA ATTCA TT | 38 | 0 | 0 | 9 | 94 | 144190806 | TATAAA GATCA TATTC CT** | 0 | 0 | 20 | >100 |
| 85 | 144154854 | TATAAA TCATG CTGCT GT* | 15 | 5 | 0 | 12 | 95 | 144195568 | TATAAA GATCA TATTC CT** | 0 | 0 | 20 | >100 |
| 86 | 144155006 | TATAAA AAATG ATGAG TT | 0 | 0 | 0 | 29 | 96 | 144200330 | TATAAA GATCA TATTC CT** | 0 | 0 | 20 | >100 |
| 87 | 144162225 | TATAAA GATCA TATTC CT** | 0 | 0 | 20 | >100 | 97 | 144203523 | TATAAA GATCA TATTC CT** | 0 | 0 | 20 | >100 |
| 88 | 144166985 | TATAAA GATCA TATTC CT** | 0 | 0 | 20 | >100 | 98 | 144208285 | TATAAA GATCA TATTC CT** | 0 | 0 | 20 | >100 |
| 89 | 144171725 | TATAAA GATCA TATTC CT** | 0 | 0 | 20 | >100 | 99 | 144211470 | TATAAA GATCA TATTC CT** | 0 | 0 | 20 | >100 |
| 90 | 144174916 | TATAAA GATCA TATTC CT** | 0 | 0 | 20 | >100 | 100 | 144214653 | TATAAA GATCA TATTC CT** | 0 | 0 | 20 | >100 |
| 91 | 144178099 | TATAAA GATCA TATTC CT** | 0 | 0 | 20 | >100 | 101 | 144217836 | TATAAA GATCA TATTC CT** | 0 | 0 | 20 | >100 |
| 92 | 144181288 | TATAAA GATCA TATTC CT** | 0 | 0 | 20 | >100 | 102 | 144222603 | TATAAA GATCA TATTC CT** | 0 | 0 | 20 | >100 |

***Contig 24.*** In DNA from 144672414 bp to 144710724 bp; 22 TATAAA motifs; IG-marker o.c.'s total: 858 (850 for $v$=1 and 8 for $v$=2).

| $t$ | $w_t$ (bp) | TATAAA[$w_t$] XXXXX XXXXX | $v$=1 | $v$=2 | $v$=16 | $s^*_{min}$ | $t$ | $w_t$ (bp) | TATAAA[$w_t$] XXXXX XXXXX | $v$=1 | $v$=2 | $v$=16 | $s^*_{min}$ |
|---|---|---|---|---|---|---|---|---|---|---|---|---|---|
| 103 | 144673296 | TATAAA TCATG CTGCT AT* | 26 | 5 | 0 | 11 | 114 | 144685302 | TATAAA ATAGG GGTTG CA | 60 | 0 | 0 | 10 |
| 104 | 144673311 | TATAAA GACAC ATGCA CA | 0 | 0 | 0 | 14 | 115 | 144688514 | TATAAA AGTTC TTGGC TC | 56 | 0 | 0 | 10 |
| 105 | 144674946 | TATAAA AGAAA CAATA CA | 31 | 0 | 0 | 10 | 116 | 144689618 | TATAAA ACAGG GAGAA GC | 23 | 0 | 0 | 11 |
| 106 | 144676088 | TATAAA CAACG GAAAT GT | 94 | 0 | 0 | 9 | 117 | 144690874 | TATAAA TAGCA GCAGA TA | 53 | 0 | 0 | 12 |
| 107 | 144676386 | TATAAA CTGGT GGGGG AG | 28 | 0 | 0 | 10 | 118 | 144691046 | TATAAA TCTAC ATCTA TG | 53 | 0 | 0 | 9 |
| 108 | 144679085 | TATAAA TATAT ATAAA GA* | 20 | 3 | 0 | 11 | 119 | 144691216 | TATAAA AATGG AGATA TT | 40 | 0 | 0 | 11 |
| 109 | 144679095 | TATAAA GAAAA ATATC AA | 20 | 0 | 0 | 10 | 120 | 144692432 | TATAAA AACTA TATTA AT | 53 | 0 | 0 | 10 |
| 110 | 144680460 | TATAAA TTAGG TTTGC TT | 38 | 0 | 0 | 9 | 121 | 144701768 | TATAAA TCTCA CAGGA CC | 67 | 0 | 0 | 11 |
| 111 | 144682953 | TATAAA TCAGA AAAGT CT | 35 | 0 | 0 | 10 | 122 | 144701786 | TATAAA ACAAA AATAC AA | 0 | 0 | 0 | 13 |
| 112 | 144683130 | TATAAA CTGAT CAACA GA | 46 | 0 | 0 | 9 | 123 | 144702545 | TATAAA ACCAT CACTA AT | 59 | 0 | 0 | 10 |
| 113 | 144684141 | TATAAA AACAA TAACA AT | 0 | 0 | 0 | 13 | 124 | 144705653 | TATAAA AATTC TAGAA GG | 48 | 0 | 0 | 10 |

***Contig 27.*** In DNA from 146214651 bp to 146253299 bp; 4 TATAAA motifs; IG-marker o.c.'s total: 150 for $v$=2.

| $t$ | $w_t$ (bp) | TATAAA[$w_t$] XXXXX XXXXX | $v$=1 | $v$=2 | $v$=16 | $s^*_{min}$ | $t$ | $w_t$ (bp) | TATAAA[$w_t$] XXXXX XXXXX | $v$=1 | $v$=2 | $v$=16 | $s^*_{min}$ |
|---|---|---|---|---|---|---|---|---|---|---|---|---|---|
| 125 | 146244959 | TATAAA GCAAG GCTTG GC* | 0 | 42 | 0 | 12 | 127 | 146251241 | TATAAA GCAAG GCTTG GC* | 0 | 42 | 0 | 12 |
| 126 | 146246826 | TATAAA TACAA AATGT TC* | 0 | 33 | 0 | 10 | 128 | 146253117 | TATAAA TACAA AATGT TC* | 0 | 33 | 0 | 10 |

Column '$t$' : ordinal number of TATAAA[$w_t$]. Column '$w_t$ (bp)': TATAAA[$w_t$] position in the DNA sequence.
Column 'TATAAA[$w_t$]XXXXX XXXXX XX': TATAAA[$w_t$] and sequence $\Im_s(w_t)$ for $s$ = 12 (see (14)).
Column '$v$ = 1': number of IG-marker o.c.'s in $\Im_s(w_t)$, $s$ = 12, for $v$ = 1. Column '$v$ = 2': the same for $v$ = 2. Column '$v$ = 16': the same for $v$ = 16. Column '$s^*_{min}$': minimal length of unique sequence after TATAAA[$w_t$] being a $G^*$-marker for single TATAAA[$w_t$] having volume $v$ = 1.
In sequences non-marked by * or **, bases shown in red represent examples of IG-marker o.c. for TATAAA[$w_t$] group with $v$ = 1. In sequences marked by *, bases shown in red represent examples of IG-marker o.c. for TATAAA[$w_t$] group with $v$ = 2. In sequences marked by **, bases shown in red represent examples of IG-marker o.c. for TATAAA[$w_t$] group with $v$ = 16.
Sequences highlighted in gray show TATAAA motifs without IG-markers (in all these cases, $s^*_{min}$ > 12).

### 3.2 Types of TATAAA inclusion into TATAAA motif groups

It follows from equation (7) that IG-markers for any given TATAAA motif provide the inclusion of the TATAAA motif into one or more TATAAA motif groups simultaneously.

Suppose *B* is any given TATAAA motif. Five types of *B* inclusion into TATAAA groups were found in DB1:

- Type 1. *B* itself forms only one group with volume $v$ = 1 (*B* is "included into itself" only).
- Type 2. *B* is included simultaneously into the two groups. One group consist of only *B* itself ($v$ = 1). Another group has volume $v$ = 2. This group consists of B and another TATAAA motif.
- Type 3. *B* is included only into the TATAAA group with volume $v$ = 2.
- Type 4. *B* is included only into the TATAAA group with volume $v$ = 16.
- Type 5. *B* is not included in any group of TATAAA since there are no IG-markers for *B*.

Table 4 shows the distribution these five types of *B* inclusion into TATAAA groups (for DB1).

**Table 4.** The distribution of TATAAA motifs representing different types of inclusion into TATAAA groups in DB1.

| Types of TATAAA motif inclusion into TATAAA motif groups | Contig regions | | | | | Total of TATAAA motifs |
|---|---|---|---|---|---|---|
| | 2 | 7 | 21 | 24 | 27 | |
| Type 1: into one TATAAA group with $v$ = 1 | 39 | 26 | 2 | 17 | 0 | 84 |
| Type 2: into two TATAAA groups with $v$ = 1 and $v$ = 2 | 1 | 4 | 1 | 2 | 0 | 8 |
| Type 3: into one TATAAA group with $v$ = 2 | 0 | 2 | 0 | 0 | 4 | 6 |
| Type 4: into one TATAAA group with $v$ = 16 | 0 | 0 | 16 | 0 | 0 | 16 |
| Type 5: into empty TATAAA group; no IG-markers | 4 | 6 | 1 | 3 | 0 | 14 |
| Total of TATAAA motifs: | 44 | 38 | 20 | 22 | 4 | 128 |

For further information on types 2 and 3 see Table 5.

As mentioned above, the inclusion of TATAAA motif into one or more TATAAA groups is governed by IG-markers. Thus, relations of TATAAA groups arise. As an example, Table 5 shows participation of IG-markers in the relations of TATAAA groups with volume $v$ = 2 and $v$ = 1 if TATAAA motifs with type 2 or 3 inclusion into TATAAA groups take place (see Table 4).

**Table 5.** The TATAAA motifs with type 2 and 3 of inclusions in TATAAA motif groups.

| (1) | (2)<br>'$v$ = 1' | (3)<br>left-hand TATAAA motif | (4)<br>'$v$ = 2' | (5)<br>right-hand TATAAA motif | (6)<br>'$v$ = 1' |
|---|---|---|---|---|---|
| 1 | 62 | TATAAA[$w_t$]; $t$ = 12, contig 2 | 1 | TATAAA[$w_t$]; $t$ = 54, contig 7 | 42 |
| 2 | 34 | TATAAA[$w_t$]; $t$ = 48, contig 7 | 19 | TATAAA[$w_t$]; $t$ = 75, contig 7 | 35 |
| 3 | 0 | TATAAA[$w_t$]; $t$ = 69, contig 7 | 12 | TATAAA[$w_t$]; $t$ = 76, contig 7 | 0 |
| 4 | 45 | TATAAA[$w_t$]; $t$ = 78, contig 7 | 3 | TATAAA[$w_t$]; $t$ = 108, contig 24 | 20 |
| 5 | 15 | TATAAA[$w_t$]; $t$ = 85, contig 21 | 5 | TATAAA[$w_t$]; $t$ = 103, contig 24 | 26 |
| 6 | 0 | TATAAA[$w_t$]; $t$ =125, contig 27 | 42 | TATAAA[$w_t$]; $t$ = 127, contig 27 | 0 |
| 7 | 0 | TATAAA[$w_t$]; $t$ =126, contig 27 | 33 | TATAAA[$w_t$]; $t$ = 128, contig 27 | 0 |

Here, Columns (3) and (5) shows 14 TATAAA motifs. For every TATAAA[$w_t$], index $t$ and the ordinal number of contig are shown. For any specific $t$, position $w_t$ is listed in Table 3. These 14 TATAAA motifs form 7 TATAAA groups with volume $v$ = 2. In DB1, these groups represent all groups with volume $v$ = 2 (see Table 4, Type 2 and 3). Ordinal numbers of groups (from 1 to 7) are shown in Column (1). In any row, a couple of TATAAA represents one group. Column (4) shows the numbers of IG-markers that are DNA identifiers (DNA landmarks) for TATAAA group with $v$ = 2. For instance, the TATAAA group No 1 has 1 IG-marker as identifier; the group No 2 has 19 IG-markers as identifiers, and so on. Every IG-marker that identifies the group with volume $v$ = 2 is represented by two IG-marker o.c.'s (see (13) when $v$ = 2).

Now let us consider Columns (2) and (6). Column (2) shows the numbers of IG-markers that are identifiers (landmarks) for any left-hand TATAAA group from Column (3) as a unique group with volume $v$ = 1. Column (6) shows the numbers of IG-markers that are identifiers for any right-hand TATAAA from Column (5). For instance, in the group No 1, left-hand TATAAA group is identified by 62 IG-markers as the unique group with volume $v$ = 1; right-hand TATAAA group is identified by 42 IG-markers as another unique group with volume $v$ = 1.

Taking into account the above-mentioned, we see that 8 (of 14) TATAAA marked in gray form 4 groups (No's 1, 2, 4 and 5) with volume $v$ = 2 and simultaneously the same 8 TATAAA form 8 groups with volume $v$ = 1 (see also Table 4, row for type 2).

On the other hand, 6 (of 14) TATAAA (non-marked in gray) form *the only groups with* $v$ = 2 (No's 3, 6 and 7) and each of the same 6 TATAAA *does not form the group with volume* $v$ = 1. It is interesting and important for us that the IG-markers that provide the inclusion of TATAAA into different groups may have the location in the same small site of DNA (12 bp in our specific case).

*Thus, we may conclude that IG-markers are the flexible tools for forming TATAAA groups and access to them.*

To make it clearer, the most detailed information on all IG-markers (and their o.c.'s), that govern the couple of TATAAA motifs in the group No 5 is presented in Table 6. 5 IG-markers are identifiers for the mentioned TATAAA group. In DNA, every one of these IG-markers is represented by two IG-marker o.c.'s. There are 15 IG-markers that identify the left-hand TATAAA motif as a unique group with volume $v$ = 1. At the same time, 26 IG-markers identify the right-hand TATAAA motif as another unique group with volume $v$ = 1. Together with their localization in DNA, all these 15, 26 and 5 IG-markers are shown in Table 6.

**Table 6.** Sets A, B and C of IG-marker o.c.'s within the sequences $\Im_s(w_t)$, $s$ = 12, downstream of the TATAAA[$w_t$], $t$ = 85 and $t$ = 103 (both TATAAA motifs belong to the TATAAA motif group of number 5 presented in Table 5).

**A.** Set of 15 IG-marker o.c.'s for single TATAAA[$w_t$ = 144154854], t = 85, contig 21; $v$ = 1: 1 IG-marker = 1 IG-marker o.c.

| | 1 | 2 | 3 | 4 | 5 | 6 | 7 | 8 | 9 | 10 | 11 | 12 | $\overline{\Delta P}$ | $r$ | $g$ | $l$ |
|---|---|---|---|---|---|---|---|---|---|---|---|---|---|---|---|---|
| IG-marker1 | T | C | A | T | G | C | T | G | C | T | G | T | 0.92 | 8 | 4 | 12 |
| ΔP | 0.96 | 0.89 | 0.83 | 0 | 0.93 | 0.97 | 0 | 0 | 0 | 0.92 | 0.95 | 0.94 | | | | |
| IG-marker2 | T | C | A | T | G | C | T | G | C | T | G | T | 0.93 | 8 | 4 | 12 |
| ΔP | 0.92 | 0.95 | 0 | 0.89 | 0.96 | 0.94 | 0.93 | 0 | 0 | 0 | 0.88 | 0.98 | | | | |
| IG-marker3 | T | C | A | T | G | C | T | G | C | T | G | T | 0.92 | 8 | 4 | 12 |
| ΔP | 0.88 | 0.88 | 0 | 0.89 | 0.89 | 0.91 | 0 | 0.93 | 0 | 0 | 0.96 | 0.98 | | | | |
| IG-marker4 | T | C | A | T | G | C | T | G | C | T | G | T | 0.90 | 8 | 4 | 12 |
| ΔP | 0.92 | 0.94 | 0 | 0 | 0.96 | 0.93 | 0.83 | 0 | 0 | 0.89 | 0.89 | 0.86 | | | | |
| IG-marker5 | T | C | A | T | G | C | T | G | C | T | G | T | 0.87 | 8 | 4 | 12 |
| ΔP | 0.90 | 0.88 | 0 | 0 | 0.90 | 0.90 | 0 | 0.90 | 0.89 | 0 | 0.85 | 0.75 | | | | |
| IG-marker6 | T | C | A | T | G | C | T | G | C | T | G | T | 0.89 | 8 | 4 | 12 |
| ΔP | 0.96 | 0.83 | 0 | 0 | 0.95 | 0.93 | 0 | 0.83 | 0 | 0.89 | 0.86 | 0.89 | | | | |
| IG-marker7 | T | C | A | T | G | C | T | G | C | T | G | T | 0.91 | 8 | 4 | 12 |
| ΔP | 0.91 | 0 | 0.88 | 0.93 | 0.92 | 0.94 | 0 | 0.91 | 0 | 0 | 0.78 | 0.98 | | | | |
| IG-marker8 | T | C | A | T | G | C | T | G | C | T | G | T | 0.91 | 8 | 4 | 12 |
| ΔP | 0.92 | 0 | 0 | 0.83 | 0.86 | 0.95 | 0 | 0.94 | 0 | 0.88 | 0.96 | 0.95 | | | | |
| IG-marker9 | T | C | A | T | G | C | T | G | C | T | G | T | 0.82 | 9 | 3 | 12 |
| ΔP | 0.89 | 0.91 | 0.86 | 0.75 | 0.80 | 0 | 0.80 | 0 | 0 | 0.67 | 0.82 | 0.90 | | | | |
| IG-marker10 | T | C | A | T | G | C | T | G | C | T | G | T | 0.76 | 9 | 3 | 12 |
| ΔP | 0.90 | 0.75 | 0.80 | 0 | 0.75 | 0 | 0.75 | 0.86 | 0.80 | 0 | 0.75 | 0.50 | | | | |
| IG-marker11 | T | C | A | T | G | C | T | G | C | T | G | T | 0.82 | 9 | 3 | 12 |
| ΔP | 0.90 | 0.80 | 0.75 | 0 | 0.80 | 0 | 0.88 | 0.75 | 0 | 0.80 | 0.85 | 0.88 | | | | |
| IG-marker12 | T | C | A | T | G | C | T | G | C | T | G | T | 0.79 | 9 | 3 | 12 |
| ΔP | 0.88 | 0 | 0.86 | 0.75 | 0.86 | 0 | 0.67 | 0.83 | 0.80 | 0 | 0.67 | 0.83 | | | | |
| IG-marker13 | T | C | A | T | G | C | T | G | C | T | G | T | 0.75 | 9 | 3 | 12 |
| ΔP | 0.80 | 0 | 0.75 | 0 | 0.90 | 0.75 | 0 | 0.67 | 0.67 | 0.75 | 0.71 | 0.75 | | | | |
| IG-marker14 | T | C | A | T | G | C | T | G | C | T | G | T | 0.82 | 9 | 3 | 12 |
| ΔP | 0.86 | 0 | 0.83 | 0 | 0.97 | 0 | 0.75 | 0.88 | 0.80 | 0.75 | 0.75 | 0.80 | | | | |
| IG-marker15 | T | C | A | T | G | C | T | G | C | T | G | T | 0.76 | 9 | 3 | 12 |
| ΔP | 0.92 | 0 | 0 | 0 | 0.75 | 0.83 | 0.75 | 0.86 | 0.50 | 0.75 | 0.83 | 0.67 | | | | |

**B.** Set of 26 IG-marker o.c.'s for single TATAAA[$w_t$ = 144673296], $t$ = 103, contig 24; $v$ = 1: 1 IG-marker = 1 IG-marker o.c.

| | 1 | 2 | 3 | 4 | 5 | 6 | 7 | 8 | 9 | 10 | 11 | 12 | $\overline{\Delta P}$ | $r$ | $g$ | $l$ |
|---|---|---|---|---|---|---|---|---|---|---|---|---|---|---|---|---|
| IG-marker1 | T | C | A | T | G | C | T | G | C | T | A | T | 0.97 | 7 | 3 | 10 |
| ΔP | 0 | 0.98 | 0 | 0 | 0.99 | 0.96 | 0.94 | 0.98 | 0.98 | 0 | 0.96 | 0 | | | | |
| IG-marker2 | T | C | A | T | G | C | T | G | C | T | A | T | 0.92 | 8 | 4 | 12 |
| ΔP | 0.88 | 0.92 | 0 | 0 | 0.93 | 0.94 | 0.96 | 0 | 0.89 | 0 | 0.89 | 0.96 | | | | |
| IG-marker3 | T | C | A | T | G | C | T | G | C | T | A | T | 0.89 | 8 | 4 | 12 |
| ΔP | 0.90 | 0.90 | 0 | 0 | 0.95 | 0.89 | 0.80 | 0 | 0 | 0.89 | 0.89 | 0.92 | | | | |
| IG-marker4 | T | C | A | T | G | C | T | G | C | T | A | T | 0.72 | 9 | 2 | 11 |
| ΔP | 0.67 | 0.67 | 0.80 | 0.50 | 0.90 | 0.80 | 0 | 0.86 | 0.75 | 0 | 0.50 | 0 | | | | |
| IG-marker5 | T | C | A | T | G | C | T | G | C | T | A | T | 0.70 | 9 | 2 | 11 |
| ΔP | 0.50 | 0.50 | 0.67 | 0.50 | 0.86 | 0.86 | 0 | 0.96 | 0 | 0.75 | 0.67 | 0 | | | | |
| IG-marker6 | T | C | A | T | G | C | T | G | C | T | A | T | 0.72 | 9 | 2 | 11 |
| ΔP | 0.67 | 0.67 | 0.50 | 0 | 0.88 | 0.86 | 0.50 | 0.67 | 0 | 0.83 | 0.87 | 0 | | | | |
| IG-marker7 | T | C | A | T | G | C | T | G | C | T | A | T | 0.77 | 9 | 2 | 11 |
| ΔP | 0.83 | 0.75 | 0.50 | 0 | 0.86 | 0.86 | 0.75 | 0 | 0.67 | 0.80 | 0.88 | 0 | | | | |
| IG-marker8 | T | C | A | T | G | C | T | G | C | T | A | T | 0.64 | 9 | 2 | 11 |
| ΔP | 0.50 | 0.75 | 0.67 | 0 | 0.75 | 0.75 | 0 | 0.75 | 0.50 | 0.50 | 0.60 | 0 | | | | |
| IG-marker9 | T | C | A | T | G | C | T | G | C | T | A | T | 0.68 | 9 | 3 | 12 |
| ΔP | 0.67 | 0.75 | 0.67 | 0 | 0.86 | 0.75 | 0 | 0.50 | 0 | 0.67 | 0.78 | 0.50 | | | | |
| IG-marker10 | T | C | A | T | G | C | T | G | C | T | A | T | 0.70 | 9 | 3 | 12 |
| ΔP | 0.67 | 0.80 | 0.75 | 0 | 0.75 | 0.75 | 0 | 0 | 0.50 | 0.67 | 0.67 | 0.75 | | | | |
| IG-marker11 | T | C | A | T | G | C | T | G | C | T | A | T | 0.71 | 9 | 2 | 11 |
| ΔP | 0.50 | 0.75 | 0 | 0.50 | 0.94 | 0.86 | 0.67 | 0.75 | 0 | 0.67 | 0.71 | 0 | | | | |
| IG-marker12 | T | C | A | T | G | C | T | G | C | T | A | T | 0.71 | 9 | 2 | 11 |
| ΔP | 0.50 | 0.67 | 0 | 0.50 | 0.80 | 0.75 | 0.80 | 0 | 0.75 | 0.95 | 0.67 | 0 | | | | |
| IG-marker13 | T | C | A | T | G | C | T | G | C | T | A | T | 0.70 | 9 | 2 | 11 |
| ΔP | 0.50 | 0 | 0.75 | 0.67 | 0.80 | 0.80 | 0.50 | 0.67 | 0 | 0.90 | 0.71 | 0 | | | | |
| IG-marker14 | T | C | A | T | G | C | T | G | C | T | A | T | 0.74 | 9 | 2 | 11 |
| ΔP | 0.67 | 0 | 0.67 | 0.75 | 0.75 | 0.86 | 0.80 | 0 | 0.67 | 0.80 | 0.67 | 0 | | | | |
| IG-marker15 | T | C | A | T | G | C | T | G | C | T | A | T | 0.74 | 9 | 3 | 12 |
| ΔP | 0.67 | 0 | 0.86 | 0.75 | 0.88 | 0.80 | 0.50 | 0 | 0.83 | 0 | 0.60 | 0.80 | | | | |
| IG-marker16 | T | C | A | T | G | C | T | G | C | T | A | T | 0.68 | 9 | 3 | 12 |
| ΔP | 0.75 | 0 | 0.83 | 0.75 | 0.83 | 0.83 | 0 | 0.50 | 0.67 | 0 | 0.33 | 0.67 | | | | |
| IG-marker17 | T | C | A | T | G | C | T | G | C | T | A | T | 0.64 | 9 | 3 | 12 |
| ΔP | 0.67 | 0 | 0.67 | 0.75 | 0.83 | 0.67 | 0 | 0.67 | 0 | 0.67 | 0.33 | 0.50 | | | | |
| IG-marker18 | T | C | A | T | G | C | T | G | C | T | A | T | 0.73 | 9 | 3 | 12 |
| ΔP | 0.80 | 0 | 0.67 | 0.80 | 0.75 | 0.95 | 0 | 0 | 0.67 | 0.50 | 0.67 | 0.80 | | | | |
| IG-marker19 | T | C | A | T | G | C | T | G | C | T | A | T | 0.76 | 9 | 3 | 12 |
| ΔP | 0.67 | 0 | 0 | 0.80 | 0.95 | 0.80 | 0.67 | 0.75 | 0 | 0.86 | 0.60 | 0.75 | | | | |
| IG-marker20 | T | C | A | T | G | C | T | G | C | T | A | T | 0.74 | 9 | 3 | 12 |
| ΔP | 0.50 | 0 | 0 | 0.80 | 0.83 | 0.80 | 0.67 | 0 | 0.75 | 0.86 | 0.78 | 0.67 | | | | |
| IG-marker21 | T | C | A | T | G | C | T | G | C | T | A | T | 0.70 | 9 | 1 | 10 |
| ΔP | 0 | 0.67 | 0.80 | 0.50 | 0.80 | 0.80 | 0 | 0.86 | 0.50 | 0.67 | 0.71 | 0 | | | | |
| IG-marker22 | T | C | A | T | G | C | T | G | C | T | A | T | 0.74 | 9 | 2 | 11 |
| ΔP | 0 | 0.75 | 0.83 | 0.67 | 0.83 | 0.75 | 0 | 0.86 | 0.67 | 0 | 0.67 | 0.67 | | | | |
| IG-marker23 | T | C | A | T | G | C | T | G | C | T | A | T | 0.69 | 9 | 2 | 11 |
| ΔP | 0 | 0.83 | 0.75 | 0 | 0.67 | 0.75 | 0 | 0.67 | 0.67 | 0.67 | 0.71 | 0.50 | | | | |
| IG-marker24 | T | C | A | T | G | C | T | G | C | T | A | T | 0.72 | 9 | 0 | 9 |
| ΔP | 0 | 0 | 0.75 | 0.80 | 0.80 | 0.83 | 0.67 | 0.67 | 0.50 | 0.67 | 0.82 | 0 | | | | |
| IG-marker25 | T | C | A | T | G | C | T | G | C | T | A | T | 0.76 | 9 | 1 | 10 |
| ΔP | 0 | 0 | 0.50 | 0.80 | 0.86 | 0.90 | 0.75 | 0.67 | 0.89 | 0 | 0.82 | 0.67 | | | | |
| IG-marker26 | T | C | A | T | G | C | T | G | C | T | A | T | 0.72 | 9 | 0 | 9 |
| ΔP | 0 | 0 | 0 | 0.86 | 0.83 | 0.75 | 0.80 | 0.50 | 0.67 | 0.50 | 0.80 | 0.75 | | | | |

**C.** Set of 5 IG-marker o.c.'s for both TATAAA[$w_t$ = 144154854], $t$ = 85, contig 21; TATAAA[$w_t$ = 144673296], $t$ = 103, contig 24; $v$ = 2: 1 IG-marker] = 2 IG-marker o.c.'s

| | 1 | 2 | 3 | 4 | 5 | 6 | 7 | 8 | 9 | 10 | 11 | 12 | $\overline{\Delta P}$ | $r$ | $g$ | $l$ |
|---|---|---|---|---|---|---|---|---|---|---|---|---|---|---|---|---|
| IG-marker1 | T | C | A | T | G | C | T | G | C | T | G/A | T | 6 | 9 | 3 | 12 |
| ΔP | 0.67 | 0.33 | 0.67 | 0.60 | 0.67 | 0.67 | 0 | 0.67 | 0.33 | 0 | 0 | 0.50 | | | | |
| IG-marker2 | T | C | A | T | G | C | T | G | C | T | G/A | T | 0.64 | 9 | 3 | 12 |
| ΔP | 0.80 | 0.33 | 0.67 | 0.78 | 0.75 | 0.67 | 0 | 0.75 | 0 | 0.33 | 0 | 0.67 | | | | |
| IG-marker3 | T | C | A | T | G | C | T | G | C | T | G/A | T | 0.68 | 9 | 3 | 12 |
| ΔP | 0.83 | 0 | 0.60 | 0.88 | 0.71 | 0.71 | 0.33 | 0.67 | 0 | 0.67 | 0 | 0.71 | | | | |
| IG-marker4 | T | C | A | T | G | C | T | G | C | T | G/A | T | 0.58 | 9 | 3 | 12 |
| ΔP | 0.75 | 0 | 0.60 | 0.71 | 0.60 | 0.67 | 0 | 0.67 | 0.33 | 0.33 | 0 | 0.60 | | | | |
| IG-marker5 | T | C | A | T | G | C | T | G | C | T | G/A | T | 0.48 | 10 | 2 | 12 |
| ΔP | 0.50 | 0.33 | A | 0.50 | 0.67 | 0.67 | 0.50 | 0.60 | 0.33 | 0.33 | 0 | 0.33 | | | | |

**A:** Set of IG-marker o.c.'s that ensure the inclusion of TATAAA[$w_t$ = 144154854] into the group with $v$ = 1. **B:** Set of IG-marker o.c.'s that ensure the inclusion of TATAAA[$w_t$ = 144673296] into the group with $v$ = 1. **C:** Set of IG-marker o.c.'s that ensure the inclusion of both TATAAA motifs into the group with $v$ = 2.

In **A**, **B** and **C** every row contains 12 bases of the DNA sequence $\Im_s(w_t)$, $s$ = 12. Red bases are bases of IG-marker o.c.'s. Below each base $X_q$ of an IG-marker, the value of the base essentiality (irreplaceability) **ΔP** ≡ **ΔP**($X_q$) (see (11)) is presented. The characteristics for each IG-marker are: $\overline{\Delta P}$, average value of base essentiality within an IG-marker; $r$, IG-marker's rank; $g$, number of internal gaps; $l$ = $r$ + $g$, IG-marker's length in DNA sequence.

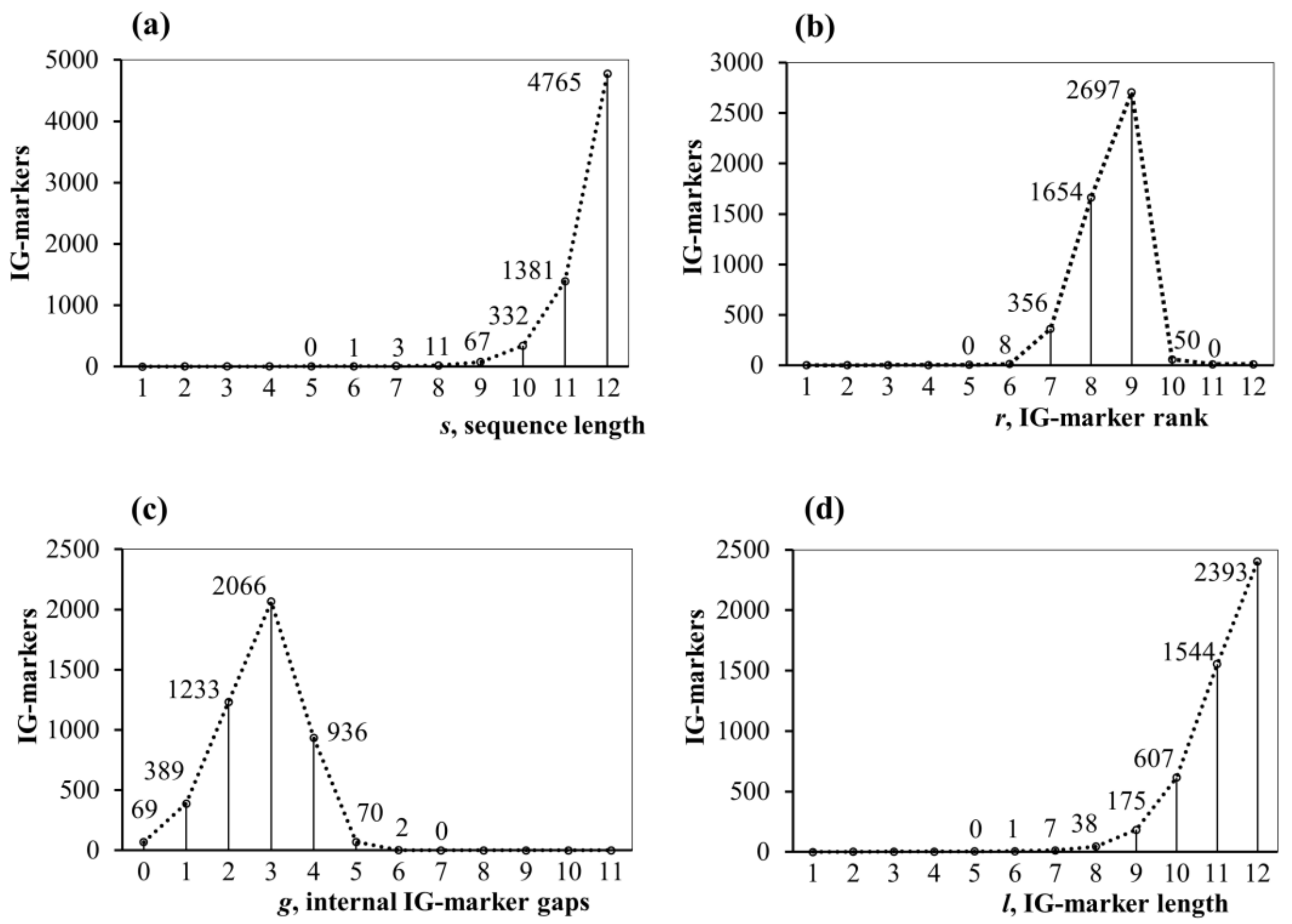

**Fig. 1.** Dependence of the total number of IG-markers on: **(a)** length *s* (bp) of sequence $\Im_s(w_t)$, $t = 1,2,\ldots,128$, $1 \le s \le 12$ (cumulative chart); **(b)** rank *r* (bp) of IG-marker; **(c)** IG-marker internal gaps *g* (bp); **(d)** IG-marker length $l = r + g$ (bp). Examples on *r*, *g*, *l* are presented in Table 6 (columns '*r*', '*g*', '*l*').

## 3.3 Set of IG-markers and parameters *s, r, g, l*

The investigated IG-markers (4765 in total) are present within sequences $\Im_s[w_t]$ of length $s = 12$ located downstream of TATAAA[$w_t$] ($t = 1,2,\ldots,128$) (see configuration (14)). Fig. 1(**a**) shows the dependency of IG-marker total number on the length *s* for $1 \le s \le 12$.

Fig. 1(**b**) illustrates the distribution of IG-markers as a function of rank value *r*. At $r = 9$, a group with maximal IG-markers is present (56.6%, 2697 of 4765). The average *r*, however, is 8.51: most IG-markers have rank 8 or 9 (91.3%, 4351 of 4765). In fact, the *r* value varies within the limits $6 \le r \le 10$. In general, the number of bases in an IG-marker cannot be too large: $r \le r_{max}$ (in our case $r_{max} = 10$), due mainly to requirement (11) that every base of an IG-marker must be essential.

Within sequences in which the IG-marker is located there are internal positions (*gaps*) occupied by bases that do not represent the IG-marker. Examples are presented in Table 6. Fig. 1(**c**) shows the distribution of the IG-markers according to number of gaps (*g*). Only 69 of the 4765 IG-markers have no gaps ($g$ = 0). Almost all of the IG-markers (98.6%, 4696 of 4765) have internal gaps, with $1 \le g \le 6$. However, the number of internal gaps for most IG-markers (88.9%; 4235 of 4765) varies between $2 \le g \le 4$ and the average $g$ is 2.76. The maximum of IG-markers number corresponds to $g$ = 3 (43.4%, 2066 of 4765).

Internal gaps ensure flexibility of a TATAAA motif groups formation. They allow different o.c.'s of any IG-marker for TATAAA motifs group having volume $v > 1$ to be localized in different DNA sequences which do not form a consensus. However, diverse IG-markers that ascertain the inclusion of individual TATAAA motifs in different TATAAA motif groups can be localized in the same area of the DNA sequence.

The length $l = r + g$ of IG-marker represents the $r$ positions of the DNA sequence that are occupied by bases making the IG-marker itself plus $g$ positions that are occupied by DNA bases of internal gaps. Fig. 1(**d**) shows the distribution of IG-markers with length $l$.

### 3.4 Potential scale of the IG-marker phenomenon

The quantity $\Re(s)$ of all potential combinatorial variants of IG-markers in a DNA fragment of length $s$ (see (14), (15)) might be estimated under the formula

$$\Re(s) = \sum_{r=1}^{r=s} 4^r \cdot \frac{s!}{r!(s-r)!} \,. \tag{19}$$

Here $r$ is the rank of the IG-marker, 4 – is the volume of the alphabet A, T, G and C. In formula (18), factor $4^r$ is absent, as (18) gives the number $\Re(s) = 2^s - 1$ of IG-marker combinatorial variants under a DNA fragment of length $s$ is fixed. We can write (19) in the form

$$\Re(s) = 4^s + \Delta(s);\ \Delta(s) = \sum_{r=1}^{r=s-1} 4^r \cdot \frac{s!}{r!(s-r)!} \,. \tag{20}$$

Summand $4^s$ is the volume of combinatorial varieties for DNA fragment of length $s$. Summand $\Delta(s)$ is the volume of combinatorial varieties in which IG-markers have internal gaps and belong to the same DNA fragment of length $s$. For $s$ = 12 we found that $6 \le r \le 10$ (see Fig. 1(**b**)). Considering this and using equation (20) we get: $4^{12}$ = 16777216; $\Delta(12)$ = 176078848. The combinatorial variety of IG-markers is ten times that of combinatory variety of DNA fragment containing IG-markers: $\Delta(12)/4^{12}$ = 10,5.

### 3.5 TATAAA motif IG-markers, summary

a) IG-markers are natural identifiers of TATAAA motifs having a minimal number of bases.

b) IG-markers ensure an unmistakable recognition of individual as well as groups of TATAAA motifs.

c) IG-markers localized in the same DNA area are capable to ensure inclusions of given TATAAA motif in several groups simultaneously (see groups 1, 2, 4 and 5 in Table 5).

d) A TATAAA motif from one group of TATAAA motifs can be separated from another TATAAA motifs belonging to the same group by a distance of millions bases in DNA (see Table 5: in the group number 1 the distance between two TATAAA motifs equal 12 885 105 bp; in the group number 4 this distance is 131 490 263 bp).

e) Almost all IG-markers contain internal gaps (see Fig. 1(**c**)).

f) The method applied here allows identification of specific IG-markers for any sequence motif on the DNA string.

Our additional data not included into this work, testify to the following:

- There are IG-markers for TATAAA motifs located in the DNA sequence sites distinct from those investigated here.
- The DNA sequence contains IG-markers not only for TATAAA motifs, but also for other motifs.
- The IG-markers phenomenon is valid for protein sequences as well.

### 3.6 IG-marker hypotheses

The following hypotheses are raised by the observed properties of IG-markers.

- *IG-markers are a part of the regulatory system of DNA functions in living cells.*
- *Promoters and introns are specific places of IG-markers location.*
- *IG-markers for TATA boxes are used by general transcription factors during initiation of transcription.*
- *GTF three-dimensional structure correlate with internal gaps of IG-markers.*
- *The minimal quantity of bases in an IG-marker ensures an energy minimum for IG-marker recognition by GTF.*

These hypotheses still require experimental verification, which we believe might be achieved through the use of molecular biology methods such as reporter gene assays and mutational approaches.

## Acknowledgements

We are grateful to Gregory Soifer for comprehensive support. We appreciate Yoram Louzoun for drawing the attention of authors to the TATA-box problem and support. The authors wish to express their thanks to Luba Trakhtenbrot, Ilya Novikov, Mark Agranovsky, Itzhak Bilkis, Sergei Sukharev and Kirill Reznik for useful discussions and assistance.